\documentclass[prl,twocolumn,showpacs]{revtex4}
\usepackage{graphicx}
\usepackage{dcolumn}
\usepackage{amsmath}
\usepackage{amssymb}
\usepackage{soul}  
\sethlcolor{yellow}
\usepackage{url}
\usepackage{hyperref}
\usepackage[dvips]{color}
\usepackage{color}
\definecolor{bluegreen}{rgb}{0,0.2,0.8}
\usepackage{subfigure,amsmath,verbatim,moreverb}
\usepackage{multirow}
\usepackage{hhline}
\def\rv{{\bf r}}
\def\uv{{\bf u}}

\begin{document}
\title{Accurate semilocal density functional for condensed matter physics and
quantum chemistry}
\author{Jianmin Tao}
\altaffiliation{Corresponding author. jianmin.tao@temple.edu} 
\affiliation{Department of Physics, Temple University, 
Philadelphia, PA 19122-1801, USA \\ Department of Chemistry, 
University of Pennsylvania, Philadelphia, PA 19104-6323, USA}
\author{Yuxiang Mo}
\affiliation{Department of Physics, Temple University, Philadelphia, 
PA 19122-1801, USA \\ Department of Chemistry, University of Pennsylvania, 
Philadelphia, PA 19104-6323, USA}

\date{\today}
\begin{abstract}
Most density functionals have been developed by 
imposing the known
exact constraints on the exchange-correlation energy, or by a fit 
to a set of properties of 
selected systems, or by both. However, accurate modeling of the 
conventional exchange hole presents a great challenge, due to the 
delocalization of the hole.
Making use of the property that the hole can be made localized under a
general coordinate transformation, here we derive an exchange hole from 
the density matrix expansion, while the correlation part is obtained 
by imposing the low-density limit constraint. From the hole, a semilocal 
exchange-correlation functional is calculated. Our comprehensive test 
shows that this functional can achieve remarkable accuracy for diverse 
properties of molecules, solids and solid surfaces, substantially 
improving upon the nonempirical functionals proposed in recent years. 
Accurate semilocal functionals based on their associated holes are 
physically appealing and practically useful for developing nonlocal 
functionals.

\pacs{31.15.E-, 31.15.ej, 71.15.Mb}
\end{abstract}

\maketitle
Kohn-Sham density functional theory (DFT)~\cite{ks65} is a mainstream
ground-state electronic structure theory, due to its high computational 
efficiency and useful accuracy. In this theory, everything is known, 
except for the exchange-correlation energy component, which has to be 
approximated as a functional of the electron density. Therefore, the 
central task of the theory is to develop consistently accurate 
exchange-correlation functional for wide-ranging problems. Many density 
functionals have been proposed~\cite{PW86,B88,LYP88,BR89,B3PW91,PBE96,
VSXC98,HCTH,PBE0,HSE03,AE05,MO6L,TPSS,revTPSS,PBEsol,SCAN15,Kaup14} and 
some of them have been widely-used in electronic structure calculations of 
molecules and solids. Most of these functionals were constructed by 
imposing exact or nearly exact energy constraints~\cite{PBE96,TPSS,
revTPSS,PBEsol}, or by a fit to a set of properties~\cite{MO6L}, or by 
their combination~\cite{VSXC98}. Because exchange and correlation parts 
have different coordinate~\cite{LP85} and spin~\cite{OP79} scaling 
properties, they are usually approximated separately.

We begin with the exchange part. For simplicity, let us consider a 
spin-unpolarized system ($n_\uparrow=n_\downarrow$), for which the 
exchange energy is defined by
\begin{eqnarray}\label{conv}
E_x = \frac{1}{2}\int d^3r \int d^3r' \frac{n(\rv)\rho_x(\rv,\rv')}
      {|\rv'-\rv|}.
\end{eqnarray}
(Atomic units $\hbar=e=m=1$ are used.) According to Eq.~(\ref{conv}), the 
exchange energy is just the electrostatic interaction of each electron at 
$\rv$ with the exchange hole at $\rv'$ surrounding the electron. The hole 
is conventionally defined as
$\rho_x(\rv,\rv') = -|\rho_1(\rv,\rv')|^2/2n(\rv)$, with
$\rho_1(\rv,\rv')=2\sum_{i=1}^{\mathrm{occup}}\psi_i^*(\rv)\psi_i(\rv')$
being the first-order reduced density matrix and $\psi_i$ being the 
occupied Kohn-Sham orbitals. We can see from Eq.~(\ref{conv}) that the
exchange energy is determined by the underlying exchange hole and the 
electron density $n(\rv)$. The exchange hole is physically meaningful.
For example, the system-averaged on-top ($|\rv'-\rv|=0$) exchange 
hole~\cite{TSP03} is proportional to the average 
electron density $\langle n \rangle$, while the latter is an experimental 
observable~\cite{ADHyman78}. In addition, the hole can be also used to 
construct higher-level nonlocal density functionals 
such as range-separation functionals~\cite{HSE03,Truhlar11}, which are 
particularly useful for the calculation of band gap and charge transfer. 
However, there is no simple procedure that can exactly extract the hole 
from a semilocal energy functional. In most cases, the hole has to be 
constructed with a reverse engineering approach~\cite{PBY96,EP98,
Henderson08,LAConstantin06}. This often introduces additional 
approximations. Therefore, it is highly desirable to approximate the 
exchange hole directly. The exchange energy functional can be easily
generated from the associated hole. 

The exchange hole can be approximated in several ways. For example, it can 
be constructed from the cutoff procedure~\cite{PW86,PBE96,PBY96}. It can be 
also constructed from simple model systems~\cite{BR89}. Here an exchange 
hole is derived from the density matrix expansion (DME) under a general 
coordinate transformation. Unlike the Taylor expansion~\cite{B83,PEZB98,
LAConstantin06}, the hole from the DME is not only correct for small 
separation (i.e., $|\rv'-\rv| \approx 0$), but also properly converged 
in the large separation limit (see discussion below). In particular, it 
automatically recovers the exact uniform-gas limit. The convergence 
property enables us to obtain the exchange energy functional, 
without resort to any numerical cutoff procedure~\cite{PW86}. Another
advantage of the DME is that the exchange hole can be made localized 
with a general coordinate transformation~\cite{TSP03}. This largely 
reduces the difficulty in the modeling of the highly nonlocal 
conventional hole. 

The DME was originally introduced by Negele and Vautherin~\cite{NV1} for the 
study of nuclear forces. Then it was generalized by Scuseria and 
co-workers~\cite{VSXC98,KOS} to calculate molecular properties, leading 
to the heavily-parametrized but accurate Voorhis-Scuseria 
functional~\cite{VSXC98}, with 21 fitting parameters. This functional 
was re-parametrized by introducing more parameters by Zhao and 
Truhlar~\cite{MO6L}, leading to MO6L, one of the most popular semilocal
functionals in quantum chemistry.

Here we introduce a novel technique in the DME. Our starting point is the 
general coordinate transformation~\cite{KOS,TSP03}
$(\rv,\rv') \rightarrow (\rv_\lambda,\uv)$, where
$\rv_\lambda = \lambda\rv + (1 - \lambda)\rv'$, $\uv=\rv'-\rv$, with
$\lambda$ being a real number between $1/2$ and 1. Since
the Jacobian of the coordinate transformation is 1, Eq.~(\ref{conv}) can
be rewritten as~\cite{TSP03}
\begin{eqnarray}\label{transform}
E_x = \frac{1}{2}\int d^3r_\lambda n(\rv_\lambda)
\int d^3u \frac{\rho_x^t(\rv_\lambda,\uv)}{u},
\end{eqnarray}
where $\rho_x^t(\rv_\lambda,\uv)$ is the transformed exchange hole 
defined by $\rho_x^t(\rv_\lambda,\uv)=|\rho_1^t(\rv_\lambda-(1-\lambda)\uv,
\rv_\lambda+\lambda\uv)|^2/2n(\rv_\lambda)$, with 
$\rho_1^t(\rv_\lambda-(1-\lambda)\uv,
\rv_\lambda+\lambda\uv)$ being the transformed 
density matrix.
 $\lambda=1$ corresponds to
the conventional hole, while $\lambda=1/2$ corresponds to the hole in the 
center of mass.

Next, we expand the transformed Kohn-Sham single-particle density matrix 
about $u=0$:
\begin{eqnarray}\label{Taylor}
\rho_1^t(\rv,\uv) =
e^{\uv\cdot[-(1-\lambda)\mbox{\boldmath$\nabla$}_1 +
\lambda \mbox{\boldmath$\nabla$}_2]}\rho_1^t(\rv,\uv)|_{u=0},
\end{eqnarray}
where $\mbox{\boldmath$\nabla$}_1$ and $\mbox{\boldmath$\nabla$}_2$ are
the gradient operators acting on the first and second arguments of the
transformed density matrix 
$\rho_1^t(\rv,\uv) = \rho_1(\rv-(1-\lambda)\uv,\rv+\lambda\uv)$, 
respectively. For convenience, the subscript $\lambda$ has been dropped 
from now on. The Taylor expansion of the density matrix can yield the 
correct small-$u$ behaviour~\cite{B83,PEZB98,LAConstantin06}, but the 
large-$u$ limit is divergent. Here we seek an expansion, which (i) 
recovers the exact uniform-gas limit, (ii) recovers the correct small-$u$ 
behaviour up to second order in $u^2$, and (iii) yields a converged 
large-$u$ limit. With these requirements, an exchange 
functional that respects the uniform-electron limit can be calculated 
from the transformed hole $\rho_x^t(\rv,\uv)$ by performing integration 
over $\uv$ in Eq.~(\ref{transform}). To achieve this goal, we introduce a 
novel {\em three-term} Bessel-function and Legendre-polynomial expansion 
of a plane wave
\begin{eqnarray}\label{planewave}
e^{x {\rm cos}\theta y}=A+B+C,
\end{eqnarray}
where
$A = \frac{1}{x}\sum_{l=0}^{\infty}(-1)^l (4l+3)j_{2l+1}(x)
Q_{2l+1}(i{\rm cos}\theta y)$,
$B = \frac{1}{x}\sum_{l=0}^{\infty}(-1)^l (4l+3)j_{2l+1}(x)
y \frac{d}{dy}Q_{2l+1}(i{\rm cos}\theta y)$,
$C = \frac{1}{x^2}\sum_{l=0}^{\infty}(-1)^l (4l+3)j_{2l+1}(x)
\frac{1}{{\rm cos}\theta}\frac{d^2}{dy^2}Q_{2l+1}(i{\rm cos}\theta y)$,
with $Q_{2l+1}(z) = P_{2l+1}(z)/z$. Eq.~(\ref{planewave}) can be derived 
with series resummation technique. (In previous works~\cite{NV1,
VSXC98,Tsuneda}, a {\em single-term} Bessel-function and 
Legendre-polynomial expansion~\cite{Arfken} was used.)  
Substituting $x=ku$ and $y=[-(1-\lambda)\mbox{\boldmath$\nabla$}_1+
\lambda\mbox{\boldmath$\nabla$}_2]/k$ into Eq.~(\ref{planewave}) and
inserting Eq.~(\ref{planewave}) into Eq.~(\ref{Taylor}) with the
transformed density matrix $\rho_1^t(\rv,\uv)$ lead to the DME 
expression
\begin{eqnarray}\label{matrix-expansion}
\rho_1^t(\rv,\uv) &=& 3n\frac{j_1(ku)}{ku}+\frac{35j_3(ku)}{2k^3u}G
+\frac{105j_3(ku)}{2k^3u^2}H,~~ 
\end{eqnarray}
where $G=\{3{\rm cos}^2\theta[(\lambda^2-\lambda+1/2)
\nabla^2 n-2\tau]+3k^2n/5\}$, 
$H={\rm cos}\theta~ (2\lambda-1)\nabla n$, with
$\tau=\sum_i^{\rm occup}|\nabla \psi_i|^2$ being the 
kinetic energy density. In the derivation of
Eq.~(\ref{matrix-expansion}), real orbitals are assumed. The first term
on the right-hand side of Eq.~(\ref{matrix-expansion}) has the form of
the density matrix of the uniform electron gas, while the second and
third terms are $\lambda$-dependent inhomogeneous corrections. Clearly, 
the general coordinate transformation only affects inhomogeneous 
corrections, but not the extended uniform electron gas, because the latter 
is translationally invariant.

To evaluate the exchange energy, we only need the spherical average of the
exchange hole over the direction of $\uv$, which is determined by the
spherical average of the square of the density matrix,
$\langle|\rho_1^t(\rv,\uv)|^2\rangle$~\cite{note1}.
In Eq.~(\ref{matrix-expansion}), there is a parameter $k$, which has the
dimension of the wave vector. $k=k_F$ is a natural choice for the uniform 
electron gas. For inhomogeneous systems, we set $k=f k_F$, where $f$ is a 
dimensionless parameter, depending on inhomogeneity~\cite{VSXC98}. If we 
choose $\lambda=1$ (conventional exchange hole), $f$ may be fixed by 
imposing the sum rule~\cite{GL76} on the model exchange hole, leading to
\begin{eqnarray}\label{sumrule}
1/f^3+70y/(9f^5)=1,
\end{eqnarray}
where $y=(2\lambda-1)^2p$ and $p=s^2=|\nabla n|^2/(2k_Fn)^2$ is the square 
of the reduced density gradient. Here we treat $\lambda$ as a free 
parameter (which will be fixed later). It was shown~\cite{JTao01,TSP03} 
that the exchange hole is not normalizable under the general coordinate 
transformation. However, as pointed out above, the general coordinate 
transformation only affects the properties of the hole for inhomogeneous 
systems. For slowly-varying densities, $1/f^3+70y/(9f^5) \approx 1$, which 
yields $f \approx 1+70y/27$. As shown by Eq.~(\ref{sumrule}), in the 
large-gradient limit, $f \rightarrow y^{1/5}$. This asymptotic behavior is 
consistent with Becke's large-gradient dependence analysis~\cite{B86}. 
Thus we assume that for any electron density,
\begin{eqnarray}
f=[1+10(70y/27)+\beta y^2]^{1/10},
\end{eqnarray}
where $\beta$ is a parameter, which will be determined together with
another parameter $\lambda$ later.

The exchange hole must be finite everywhere in space. However, the
appearance of the Laplacian of the electron density in the DME of
Eq.~(\ref{matrix-expansion}), which cannot be eliminated through the
angle average of the square of the density matrix
$\langle|\rho_1(\rv,\uv)|^2\rangle$, can make the model exchange hole 
unphysically divergent at a nucleus. Therefore, we must eliminate the 
Laplacian in Eq.~(\ref{matrix-expansion}). This can be done with the 
second-order gradient expansion of the kinetic energy density, 
$\tau=\tau^{\rm unif}+|\nabla n|^2/(72n)+\nabla^2 n/6$, where
$\tau^{\rm unif}=(3/10)k_F^2n$ is the Thomas-Fermi kinetic energy
density. This technique has been used in the development of semilocal
DFT~\cite{TPSS,revTPSS} and electron localization indicator~\cite{TLZR05}. 
Replacing the Laplacian with
$\nabla^2 n=6[\tau-\tau^{\rm unif}-|\nabla n|^2/(72n)]$ in
$\langle|\rho_1(\rv,\uv)|^2\rangle$ [or Eq.~(\ref{matrix-expansion})]
yields the spherically-averaged exchange hole 
\begin{eqnarray}\label{hole}
\rho_x^{\rm t} =
-\frac{9n}{2}\frac{j_1^2(ku)}{k^2u^2}-
\frac{105j_1(ku)j_3(ku)}{k^4u^2}L
-\frac{3675j_3^2(ku)}{8k^6u^4}M, 
\end{eqnarray}
where $L=[3(\lambda^2-\lambda+1/2)
(\tau-\tau^{\rm unif}-|\nabla n|^2/72n)-\tau+3k^2n/10]$, and
$M=(2\lambda-1)^2|\nabla n|^2/n$.
The exchange functional can be calculated from the hole by substituting
Eq.~(\ref{hole}) into Eq.~(\ref{transform}) and using the
Weber-Schafheitlin integral formula~\cite{Luke62}. The result is
\begin{eqnarray}\label{EXDME}
E_x[n] = \int d^3r~ n\epsilon^{\rm unif}(n)F_x^{\rm DME}(p,\tau),
\end{eqnarray}
where $\epsilon^{\rm unif}(n)=-3k_F/4\pi$ is the exchange energy per 
electron of the uniform electron gas, and $F_x^{\rm DME}$ is the 
enhancement factor given by $F_x^{\rm DME} = 1/f^2+7R/(9f^4)$,
with $R=1+595(2\lambda-1)^2p/54-[\tau-
3(\lambda^2-\lambda+1/2)(\tau-\tau^{\rm unif}-
|\nabla n|^2/72n)]/\tau^{\rm unif}$.

The two parameters $\lambda$ and $\beta$ can be determined by the 
following two conditions: (i) Recovery of the exchange energy of the H 
atom, and (ii) the least value that ensures $F_x^{\rm DME}$ to be a 
monotonically increasing and smooth function of the reduced density 
gradient $s$ in the iso-orbital region where 
$\tau=\tau_W=|\nabla n|^2/(8n)$. This yields $\lambda=0.6866$ and 
$\beta=79.873$. These two constraints were used in the construction of 
TPSS functional.

\begin{figure}
\includegraphics[width=\columnwidth]{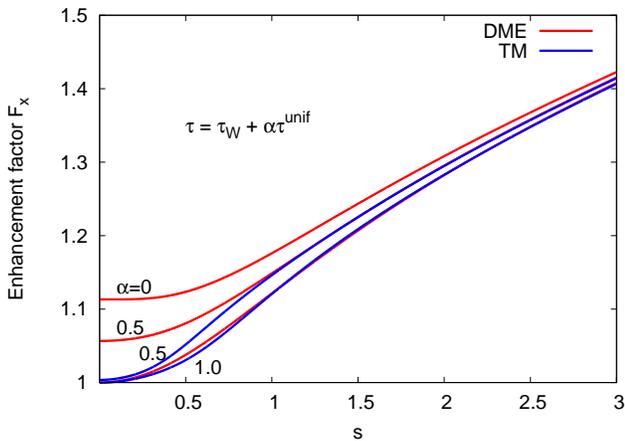}
\caption{Enhancement factors $F_x^{\rm DME}$ of Eq.~(\ref{EXDME}) 
(red) and $F_x^{\rm TM}$ of Eq.~(\ref{final}) (blue) in iso-orbital regions 
with $\alpha = 0$ and orbital overlap regions with 
$\alpha = 0.5, 1.0$. $F_x^{\rm TM} = F_x^{\rm DME}$ at $\alpha=0$.}
\label{figure1}
\end{figure}
\begin{figure}
\includegraphics[width=\columnwidth]{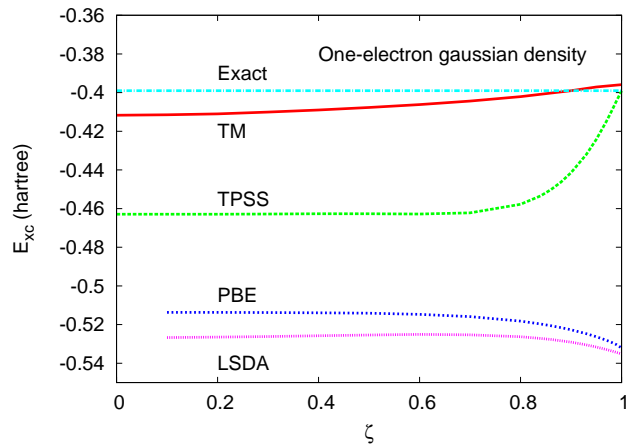}
\caption{Exchange-correlation energy in the low-density limit 
for one-electron gaussian density with constant spin 
polarization~\cite{TPSS,JCP04} 
$n_\uparrow=[(1+\zeta)/(2\pi^{3/2})]e^{-r^2}$ and
$n_\downarrow=[(1-\zeta)/(2\pi^{3/2})]e^{-r^2}$.}
\label{figure2}
\end{figure}

\begin{table*}[t]
\caption{Statistical errors of five nonempirical density functionals LSDA, 
PBE, TPSS, TMTPSS, and TM for 148 G2/97 atomization energies, 82 diatomic 
harmonic frequencies ($\omega_e$), 96 bond lengths ($r_e$), 10 
hydrogen-bond (H-bond) dissociation energies and bond lengths, 16 lattice 
constants ($a_0$) and bulk moduli ($B_0$), 7 cohesive energies 
($\epsilon_{\rm coh}$), and jellium surface exchange-correlation
energies ($\sigma_{xc}$) 
for $r_s = 2$-3. The data of the LSDA, PBE, and TPSS are taken from 
Refs.~\cite{TPSS,VNStaroverov04,JCP03}. The best values are in blue. 
ME = mean error and MAE = mean absolute error.}
\begin{tabular}{c|c|c|c|c|c|c|c|c|c|}
\hline \hline  
  &\multirow{2}{40pt}{AE6~~G2 (kcal/mol)}&
  \multirow{2}{20pt}{$\omega_e$ (cm$^{-1}$)} &
  \multirow{2}{20pt}{$r_e$ ($\AA$)} &
  \multirow{2}{40pt}{H-bond (kcal/mol)} & 
  \multirow{2}{45pt}{H-bond length ($\AA$)}&
  \multirow{2}{35pt}{$~~~\sigma_{xc}$ (erg/cm$^2$)} &
  \multirow{2}{10pt}{$a_0$ ($\AA$)} &
  \multirow{2}{20pt}{$B_0$ (GPa)} &
  \multirow{2}{20pt}{$\epsilon_{\rm coh}$ (eV)} \\ 
        &    &  & & & & & & &  \\ \hline
        &MAE~   MAE &ME~  MAE &ME~  MAE &ME~  MAE&
         ME~  MAE&ME~ MAE &ME~  MAE &ME~  MAE &ME~  MAE \\ \hline

LSDA &77.3~ 83.5     &$-11.8$~  48.9   &$0.001$~ 0.013   &5.8~ 5.8         &$-0.127$~ 0.147   &$-78$~ 78         &$-0.072$~ 0.072        &12.6~ 13.2       &0.68~  0.68  \\
PBE  &15.5~ 18.3     &$-31.7$~  42.0   &0.015~ 0.016      &0.9~ 1.0         &$-0.018$~ 0.043   &$-133$~ 133    &$-0.047$~ 0.050      &$-4.2$~ {\textcolor{blue}{\bf 5.9}}&$-0.03$~ 0.12 \\
TPSS &5.9~ 6.2 &$-18.7$~  30.4   &0.014~ 0.014      &0.3~ 0.6         &$-0.006$~ 0.021   &$-60$~ 60    &$-0.034$~ 0.036      &0.0~ 8.8         &$-0.18$~ 0.18 \\
TMTPSS&5.5~  {\textcolor{blue}{\bf 5.2}}  &$-16.6$~  {\textcolor{blue}{\bf 29.0}} &0.013~ 0.013 &$-0.5$~ 0.5 &0.036~ 0.041 &$-1$ ~ {\textcolor{blue}{\bf 1}}&0.005~  0.019 &4.9~ 7.3 &$-0.13$~ 0.13 \\ 
TM  &{\textcolor{blue}{\bf 5.1}}~  6.5&$-16.6$~ 29.7&0.010~ {\textcolor{blue}{\bf 0.012}}&$-0.1$~ {\textcolor{blue}{\bf 0.3}}&0.014~ {\textcolor{blue}{\bf 0.017}}&35~ 35&$-0.003$~ {\textcolor{blue}{\bf 0.017}}&6.3~ {\textcolor{green}{\bf 7.0}}         &$-0.05$~ {\textcolor{blue}{\bf 0.08}} \\
\hline\hline
\end{tabular}
\label{table1}
\end{table*}

The typical bulk valence electron density is slowly-varying. Recovery of 
the correct gradient expansion of the exchange energy is important for 
solids. It is also crucial for surface energy, because
it involves the bulk solid contribution. However, 
the exchange energy functional and the underlying exchange hole from the 
DME are only exact in the uniform-gas limit, but not for slowly varying 
densities. To fix this problem, we propose the following interpolation 
formula between the compact density (where the DME is more suitable) and 
the delocalized slowly-varying density:
\begin{eqnarray}
\rho_x(\rv,\uv)=w \rho_x^t(\rv,\uv)+(1-w)\rho_x^{\rm sc}(\rv,\uv),
\end{eqnarray}
with $w = [(\tau_W/\tau)^2+3(\tau_W/\tau)^3]/[1+(\tau_W/\tau)^3]^2$ being
the weight between the compact density and the slowly varying correction
(sc). (Other forms of $w$ are possible. This one provides a slightly 
more balanced interpolation between the compact density and the slowly 
varying density.) Near a bond center of molecules, $w\approx 0$ 
(except for one or two-electron systems, in which $w$ is identically 1 
everywhere). In the core region and density tail, $w\approx 1$. In bulk 
solids, $w$ is small. $\rho_x^{\rm sc}(\rv,\uv)$ can be obtained from the 
slowly varying gradient expansion of the exchange 
hole~\cite{LAConstantin06}. This yields the final expression
\begin{eqnarray}\label{final}
F_x = w F_x^{\rm DME} + (1-w) F_x^{\rm sc},
\end{eqnarray}
where $F_x^{\rm sc}$ is the fourth-order gradient correction given by
$F_x^{\rm sc} = \{1+10[(10/81+50p/729)p+146{\tilde q}^2/2025-
(73{\tilde q}/405)[3\tau_W/(5\tau)](1-\tau_W/\tau)]\}^{1/10}$, with 
$\tilde{q} = (9/20)(\alpha - 1) +  2p/3$.

This completes the spin-unpolarized case. The hole and exchange
energy functional can be easily generalized to any spin polarization,
with the spin-scaling relation~\cite{OP79}
$\rho_x[n_\uparrow,n_\downarrow] = (n_\uparrow/n)\rho_x[2n_\uparrow]+
(n_\downarrow/n)\rho_x[2n_\downarrow]$. For convenience, we
call it TM.  

The density overlap region is an important region, where the magnitude 
of the first derivative of the density is small, but higher-order 
derivatives can be large. Therefore, it is a pseudo-slowly varying region. 
It includes intershell region in atoms, multiple-bond congestion region, 
and interstitial region in metals. This region can be modeled with
$\tau = \tau_W + \alpha \tau_0$. 

Figure~\ref{figure1} shows the variations of the enhancement factor
$F_x^{\rm DME}$ [Eq.~(\ref{EXDME})] and its slowly-varying corrected 
version [Eq.~(\ref{final})] from iso-orbital ($\alpha=0$) 
to overlap regions ($\alpha >0$) in the range $0 \le s \le 3$.
In the iso-orbital region, $F_x^{\rm TM}$ reduces to $F_x^{\rm DME}$, 
while in the overlap region, $F_x^{\rm TM}$ becomes relatively de-enhanced 
at small $s$, due to the order-of-limit problem~\cite{TPSS}. Since this 
only happens near a nucleus, it is harmless. In the iso-orbital or core region, 
both enhancement factors are flat so that the exchange potential in this 
region remains finite, like LSDA and TPSS meta-GGA, but unlike GGA.

Now we turn to the correlation part. We seek for a correlation energy 
functional with the underlying correlation hole. The correlation 
functional should respect three important constraints: (i) one-electron 
self-interaction-free, (ii) correct for slowly-varying densities, and 
(iii) exact or nearly exact in the low-density or strong-interaction 
limit, in which the exchange-correlation energy is 
spin-independent~\cite{SPK}.
These considerations lead us to assume that our correlation takes the 
same form as the TPSS correlation (Eqs.~(11) and~(12) of 
Ref.~\cite{TPSS}), but replaces $C(\zeta,\xi)$ by a simpler form
\begin{eqnarray}
C(\zeta,\xi) = \frac{0.1\zeta^2 + 0.32\zeta^4}{\{1 +  \xi^2
[(1 + \zeta)^{-4/3} + (1 - \zeta)^{-4/3}]/2 \}^4}, 
\label{eq_CCC}
\end{eqnarray}
where $\zeta = (n_\uparrow - n_\downarrow)/n$, and 
$\xi = |\nabla \zeta|/2(3\pi^2n)^{1/3}$. The coefficients 0.1 and 0.32 
are obtained by keeping $E_{xc}$ for the one-electron gaussian 
density~\cite{TPSS,JCP04} as spin-independent as possible, when $\zeta$ varies 
from 0 to 1, so that constraint (iii) is well respected. As shown in 
Fig.~\ref{figure2}, this modification considerably improves the low-density 
limit of TPSS, leading to much better agreement of TM with the exact 
$E_{xc}$ of the one-electron gaussian density and smoother variation all 
the way from $\zeta = 0$ to 1. In addition, the hole underlying this 
correlation functional is known~\cite{LAConstantin06}. 
 
Finally, we make a comprehensive assessment of TM functional on 
molecules, solids, and surfaces. To do this, we implement TM into 
the Gaussian program~\cite{G09} by locally modifying the G09 code. 
Molecular test includes 
148 G2/97 atomization energies~\cite{Robin12}, 96 bond lengths ($r_e$), 
82 harmonic frequencies ($\omega_e$), and 10 h-bond dissociation energies 
and bond lengths, while solid test includes 16 lattice constants ($a_0$) 
and bulk moduli ($B_0$) as well as 7 cohesive energies 
($\epsilon_{\rm coh}$). All molecular calculations were performed 
self-consistently with basis set 6-311++G($3df,3pd$), while for 
solid-state calculations, we used the basis sets given in 
Refs.~\cite{VNStaroverov04,RKrishnan80,ADMcLean80}.
Since the RPA (random-phase approximation) surface correlation energy 
is not reliable in the low-density regime, only $\sigma_{xc}$ is reported 
for $r_s=2$ to $3$~\cite{BWood07,Lucian08}.
The results are summarized in Table~\ref{table1}. The detailed comparison 
can be found from the Supplemental Material (SM)~\cite{SM}. To show the 
trend, the relative error for each property is also given in SM. From 
Table~\ref{table1}, we see that TM yields remarkable improvement over the 
LSDA, PBE, and TPSS functionals for nearly all the properties considered.

TM is also superior, compared to other density functionals. For example, 
the error of TM on AE6 atomization energies, which are representative of 
223 G3 molecules, is only 5.1 kcal/mol, while the MAE of revTPSS on this 
special set is 5.9 kcal/mol~\cite{revTPSS} (see Table S4 for detail). 
The h-bond description of TM is much more accurate than those of both 
TPSS and revTPSS~\cite{revTPSS}.
As shown in Table S6, the error of TM for 16 lattice constants 
(MAE = 0.017 \AA) is smaller than both PBEsol (MAE = 0.021 \AA) and 
revTPSS (MAE = 0.031 \AA)~\cite{revTPSS}. From Table S7, we can see that 
the cohesive energies of TM (MAE = 0.08 eV/atom) are more accurate 
than those of revTPSS (MAE = 0.14 eV/atom)~\cite{JSUN11}. TM is 
competitive with or more accurate 
than the SCAN meta-GGA developed recently by Sun, Ruzsinszky,
and Perdew~\cite{SCAN15}, although the latter contains several empirical 
parameters fitted to atoms and a van der Waals system (Ar2 dimer). For 
example, SCAN predicts enthalpy of formation for G3/223 molecules with 
MAE = 5.7 kcal/mol, which is close to that of TPSS (MAE = 5.8 kcal/mol), 
while TM is less accurate than TPSS for G2 atomization energies by 
0.3 kcal/mol. However, the error of TM in lattice constant is smaller than 
that of SCAN (MAE = 0.019 \AA), as shown in Table S6. 

Table~\ref{table1} also shows that TM is more accurate than the 
combination, TMx+TPSSc or TMTPSS, for most properties. This demonstrates 
why the improvements in correlation from Eq.~(\ref{eq_CCC}) are 
important.

In summary, we have developed an exchange-correlation functional, which 
can achieve remarkable accuracy over wide-ranging properties.
Unlike other DFT methods, TM shows consistent improvement over the 
nonempirical DFT methods developed in recent years. This is particularly
important in electronic structure calculations of novel materials. Since
all the parameters in TM are determined by paradigm densities, rather than 
by particular systems, they are easily transferrable from one system to another.
This is a significant step toward the elimination of different functional 
for different task. 
We have applied TM functional to calculate low-lying 
atomic and molecular excitation energies within the adiabatic time-dependent 
DFT~\cite{JCP08}. The results are also remarkably 
accurate, substantially improving the adiabatic LSDA, PBE GGA, and TPSS 
functional. These highly accurate results provide compelling evidence of the 
power of TM functional. We will report these results elsewhere.
The most appealing feature of this functional is that 
it is essentially derived from or fully based on the underlying hole and 
thus has strong physical base, compared to those developed solely from 
energy constraints or fitting procedure. The hole combines the advantages 
of that based on the numerical cutoff procedure~\cite{PW86} and the 
hole based on the hydrogen atom~\cite{BR89} (which slightly violates the 
exact uniform-gas limit). The physics behind the derivation is transparent. 
The TM hole can be used to build nonlocality into the energy functional by 
developing range-separation functionals for band gap~\cite{FTran09,prb2011}, 
reaction barrier, and charge transfer calculations.

We thank Roberto Car for valuable comments and suggestions, 
Viktor N. Staroverov, Andrew M. Rappe, Guocai Tian, and Haowei Peng for 
helpful discussions, and Gustavo E. Scuseria for useful comments. This 
work was supported by NSF under Grant no.  CHE-1261918. Computational 
support was provided by the University of Pennsylvania and Temple 
University.


\begin{thebibliography}{100}
\bibitem{note1}
$\langle|\rho_1^t(\rv,\uv)|^2\rangle = |\langle\rho_1^t(\rv,\uv)\rangle|^2$,
where $\langle \cdots \rangle$ denotes the angle average, only holds for
$\lambda=1/2$. However, $|\langle\rho_1^t(\rv,\uv)\rangle|^2$ was also used
for other $\lambda$ in previous works~\cite{VSXC98,KOS}.
\end{thebibliography}
\end{document}